\documentclass[conference]{IEEEtran}
\pdfoutput=1 

\usepackage{pervasives} 

\begin{document}

\pagestyle{empty}
\thispagestyle{empty}

\title{\sys: Interactive System Configuration Repair}

\author{%
  \IEEEauthorblockN{%
    Aaron Weiss
  }
  \IEEEauthorblockA{%
    Northeastern University \\
	Boston, MA, USA 02115\\
    weiss@ccs.neu.edu
  }
  \and
  \IEEEauthorblockN{%
    Arjun Guha \qquad \qquad \qquad \qquad \qquad Yuriy Brun
  }
  \IEEEauthorblockA{%
    University of Massachusetts, Amherst \\
	Amherst, MA, USA 01003\\
    \{arjun, brun\}@cs.umass.edu
  }
}

\maketitle

\begin{abstract}
System configuration languages provide powerful abstractions that simplify
managing large-scale, networked systems. Thousands of organizations now use
configuration languages, such as Puppet. However, specifications written in
configuration languages can have bugs and the shell remains the simplest way
to debug a misconfigured system. Unfortunately, it is unsafe to use the shell
to fix problems when a system configuration language is in use: a fix applied
from the shell may cause the system to \emph{drift} from the state specified
by the configuration language. Thus, despite their advantages, configuration
languages force system administrators to give up the simplicity and
familiarity of the shell.

This paper presents a synthesis-based technique that allows administrators to
use configuration languages and the shell in harmony. Administrators
can fix errors using the shell and the technique automatically repairs the
higher-level specification written in the configuration language. The approach
(1)~produces repairs that are \emph{consistent} with the fix made using the
shell; (2)~produces repairs that are \emph{maintainable} by minimizing edits
made to the original specification; (3)~ranks and presents multiple repairs
when relevant; and (4)~supports all shells the administrator may wish to
use. We implement our technique for Puppet, a widely used system configuration
language, and evaluate it on a suite of benchmarks under 42 repair scenarios.
The top-ranked repair is selected by humans 76\% of the time and the
human-equivalent repair is ranked 1.31 on average.
\end{abstract}

\lstset{language=puppet}

\section{Introduction}

Modern computing systems are large, complex, and need to be reconfigured
frequently to address changing threats and requirements. The job of a
\emph{system administrator} is to perform these tasks. For example, if a web
server is under attack, she may reconfigure a firewall; if a new security patch
is available, she may deploy it; if an intrusion detection system is needed,
she may set it up and ensure it does not interfere with normal operations.
System administration is a difficult task and the majority of
large organizations use \emph{system configuration languages} to make the job
easier. For example, Puppet~\cite{puppet} is deployed at over 33,000 companies,
Chef~\cite{chef} has over 40 million downloads~\cite{2016-chef-usage}, and
Ansible~\cite{ansible} was quickly bought by Red Hat a few years after its
release.

Unfortunately, updating system configurations is a surprisingly
difficult task and several recent, high-profile computing failures have
been caused by configuration updates gone wrong. For example, in 2016, some
Google App Engine customers suffered a two-hour service outage due to a
configuration error that was triggered during an application server
update~\cite{2016-gae-incident}. In 2015, the New York Stock Exchange suffered
an outage that halted trading for four hours because a software update went
awry~\cite{2015-nyse-outage}. In 2010, Facebook suffered a 2.5 hour outage that
was again caused by a faulty configuration update~\cite{2010-fb-outage}. In
that incident, a system for verifying system configurations actually
exacerbated the problem.

This paper focuses on Puppet, the most widely deployed system
configuration language~\cite{weins:devops2017},
but our work generalizes to other
configuration languages (see \Cref{sec:related}).
Puppet configurations (known as \emph{manifests}) have the following key features.
First, manifests are declarative, parameterizable, and support modular composition.  For
example, Puppet has an online repository of nearly 5,000
community-supported manifests.
Second, manifests make systems reproducible.  For example, if a
new web server is needed, a system administrator can quickly set it up
if she already has a web server manifest. Finally, manifests support
centralized management. Puppet uses a client-server model, where all
manifests are maintained on a centralized server and
propagated to client machines.

Manifests may have bugs and even bug-free manifests
need to be updated to address changing requirements. However, there are many
cases where manifests make changes harder to apply than they should be.
A small change, such
as creating a new user, adding a firewall rule, or starting a
service, is easy to perform with the command-line shell, using
commands such as \verb|useradd|, \verb|iptables|, and
\verb|service|
that are familiar to administrators.  The shell also lets the administrator explore the
state of the system and, unlike a manifest, typically provides immediate feedback when the
administrator makes a mistake.  By contrast, editing a manifest is
much harder. First, in a large manifest that uses high-level,
user-defined abstractions, it can be difficult to find where and how an 
update should be made. Second, the only way to test an update is to
redeploy it, which can take anywhere from minutes to hours.  Finally, an update may have
unintended effects, especially if the update is in a function that is called
from multiple contexts.

The natural solution to this problem is to use a manifest
and the shell simultaneously. For example, a manifest could specify the
state of the machine while small updates are made using the shell.
Unfortunately, it is not safe to make
changes from the shell when a manifest is in use, because
the actual state of the system will no longer match the state specified
in the manifest\,---\,a phenomenon known as
\emph{configuration drift}. 

\myparagraph{Our approach}
We present a new approach to repairing system configurations, called
\emph{\technique}, which bridges the gap between the shell and system
configuration languages. \Technique is a program synthesis-based technique that
allows a manifest to be automatically repaired given a sequence of shell
commands to guide the desired system state. Therefore, instead of running the
risk of configuration drift, we allow system administrators to use a familiar
shell to fix errors with the confidence that the manifest will be updated
automatically. Our approach has several important properties: %
\begin{enumerate}

\item Our repair procedure is \emph{consistent}: the synthesized repair is 
  guaranteed to preserve all changes made in the shell (that are within
  the scope of the system model), so there is no configuration drift.
  
\item Our repair procedure preserves the structure and abstractions in
  the manifest.  Therefore, we synthesize \emph{maintainable} patches.
  
\item When multiple consistent repairs exist, we rank them and present
  several alternatives in a comprehensible manner.

\item Our approach places no restrictions on how the shell is
    used and works with \emph{all} existing shells.

\end{enumerate}

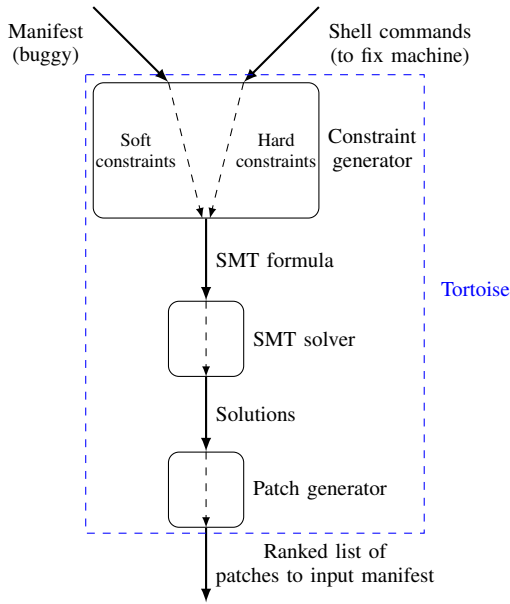
\begin{figure}
\centering
\tikzset{
  x=1cm,
  y=-1cm,
  style={font=\footnotesize,node distance=1.5cm},
  arr/.style={thick,-latex},
  node/.style={minimum width=1.6cm,minimum height=1.2cm,align=center},
  io/.style={trapezium,trapezium left angle=70, trapezium right angle=-70,minimum width=1.6cm,minimum height=1.2cm,align=center}
  }
\begin{tikzpicture}
  
  \coordinate (manifestsrc) at (1.5,0);
  \coordinate (shellsrc) at (4.5,0);
  \coordinate (manifestdst) at (2.5,1);
  \coordinate (shelldst) at (3.5,1);

  \node[rounded corners,align=center,anchor=north,draw,minimum height=1.8cm,minimum width=3cm] at (3,1) (cgen) {};

  \node[rounded corners,below of=cgen,align=center,anchor=north,draw,minimum height=1cm,minimum width=1cm,yshift=-0.5cm] (smt_solver) {};
  \node[rounded corners,below of=smt_solver,align=center,anchor=north,draw,minimum height=1cm,minimum width=1cm]  (interp) {};

  \coordinate[below of=interp,node distance=1.5 cm] (output) {};
  
  \draw[arr] (manifestsrc) -- (manifestdst) node[xshift=-0.5cm,midway,left,align=center]{Manifest\\(buggy)};
  \draw[arr] (shellsrc) -- (shelldst) node[xshift=0.5cm,midway,right,align=center]{Shell commands\\(to fix machine)};
  \draw[dashed,-latex] (manifestdst) -- ([xshift=-.05cm]cgen.south) node[midway,left,align=center,font=\scriptsize] {Soft\\constraints};
  \draw[dashed,-latex] (shelldst) -- ([xshift=.05cm]cgen.south)  node[midway,right,align=center,font=\scriptsize] {Hard\\constraints};
  \draw[arr] (cgen.south) -- (smt_solver.north) node[midway,right,align=center]{SMT formula};
  \draw[arr] (smt_solver.south) -- (interp.north) node[midway,right,align=center]{Solutions};;
  \draw[arr] (interp.south) -- (output) node[midway,right,align=center]{Ranked list of\\patches to input manifest};
  
  \node[right of=cgen,anchor=west,align=center,node distance=1.5cm]  {Constraint\\generator};  
  \node[right of=smt_solver,anchor=west,align=center,node distance=0.5cm]  {SMT solver};
  \node[right of=interp,anchor=west,align=center,node distance=0.5cm]  {Patch generator};

  \draw[dashed,-latex] (smt_solver.north) -- (smt_solver.south);
  \draw[dashed,-latex] (interp.north) -- (interp.south);

  \draw[dashed,color=blue] (1.4,0.9) rectangle (5.9,7);
  \node[color=blue,right] at (6,3.75) {\sys};

\end{tikzpicture}
\caption{The user deploys a buggy Puppet manifest and uses the shell
to fix the machine state. \sys automatically produces
a ranked list of manifest repairs that preserve the changes made in the shell.}
\label{workflow}
\end{figure}
  
We have implemented our approach in a tool
called \sys, which is outlined in \Cref{workflow} and works as follows:

\begin{enumerate}

\item Suppose the administrator needs to update a machine that is
managed by a Puppet manifest. \sys allows her to directly update the machine
using ordinary shell commands. Behind the scenes, \sys uses ptrace to record
all system calls and file system changes made from the shell.

\item When she is done, \sys builds a model of the original manifest and
 the updates from the shell in a language called \lang. The model
 treats the shell updates as hard constraints and the original manifest
 as soft constraints.

\item \sys translates the \lang model  into logical formulae for an SMT
solver (specifically, Z3-str~\cite{zheng:z3str}). These formulae produce
$\exists \forall$-queries, which we solve using \textsc{cegis}~\cite{solarlezama:sketch}.

\item \sys interprets each solution produced by the SMT solver as a patch
to the Puppet manifest, ranks the patches in an intuitive way, and presents
the most likely patches to the user.

\item Finally, the user selects the patch she wishes to apply. Although
different patches have different effects, \sys guarantees that all patches are
\emph{consistent} and preserve the changes that the user made from the shell
 in step 1.

\end{enumerate}

We evaluate \sys on an existing suite of Puppet
manifests~\cite{shambaugh:rehearsal}.  We identify a total of
\scenecount scenarios where the manifests would need
repair. However, instead of repairing the manifests directly, as a
system administrator would normally have to do, we directly update
the system using the shell and use \sys to synthesize the repair to
the manifest. The highest-rank repair \sys synthesized was the correct repair
76\% of the time, and the correct repair was in the top five \sys-synthesized
repairs 100\% of the time. \sys and our benchmarks are available at
\url{http://plasma.cs.umass.edu/tortoise}.

The rest of this paper is organized as follows. \Cref{motivation},
motivates our approach with an example.
\Cref{repair-space} details \sys's expressiveness.  
\Cref{sec:semantics} presents the \lang language
and describes the translation from Puppet manifests and system call traces
to \lang.
\Cref{formula-generation} describes how \sys converts \lang constraints to logical formulae
for an SMT solver, how models returned by the solver are interpreted as repairs,
and how these repairs are ranked.
\Cref{sec:evaluation} evaluates \sys and
\Cref{sec:limitations} summarizes our work's limitations.
Finally, \Cref{sec:related} discusses related work and
\Cref{sec:contributions} concludes.

\section{A Configuration Repair Scenario}
\label{motivation}

\begin{figure}[t]

\begin{centerbox}
\begin{puppetcode}
package{"apache2": ensure => present }
    
service{"apache2": ensure => running }

file{"/etc/apache2/sites-enabled/piedpiper.conf":
  content => "<VirtualHost www.piedpiper.com:80>
  DocumentRoot /var/sites/piedpiper
  </VirtualHost>" }

file{"/var/sites/piedpiper:
  ensure => "directory",
  source => "puppet://sites/piedpiper",
  owner => root,
  mode => 0700, #\label{line:permissions}#
  recurse => "remote" }

Package["apache2"]
  -> Service["apache2"]
Package["apache2"]
  -> File["/etc/apache2/sites-enabled/piedpiper.conf"]
File["/etc/apache2/sites-enabled/piedpiper.conf"]
  ~> Service["apache2"]
\end{puppetcode}
\end{centerbox}
\caption{Managing a single website.}
\label{example1}
\end{figure}

To motivate the need for \technique, we first present a simple manifest that sets up a
web server (\Cref{example1}). This manifest has a bug and we describe how
a system administrator would find and fix the bug using the shell, and
then how \sys automatically synthesizes the repair to the manifest by observing the
administrator's shell commands.

A manifest is a declarative specification of the expected system
configuration. A manifest specifies a set of resources, their state, and
their interdependencies. Each resource consists of (1)~a \emph{type}, for
example \lstinline|package|, \lstinline|file|, or \lstinline|service|; 
(2)~a \emph{title}, interpreted based on the type, for example, the title of
a \lstinline|file| is the path of the file, whereas the title of a
\lstinline|package| is the name of the package; and (3)~a dictionary of
\emph{attributes} to configure the resource, such as specifying that a
\lstinline|package| should be present or absent, or that a
\lstinline|file| refers to a directory.

\Cref{example1} shows a manifest with four resources:
(1)~the \lstinline|apache2| package, which must be present on the system;
(2)~the \lstinline|apache2| service, which must be running; 
(3)~the file \lstinline|piedpiper.conf| that sets up \url{www.piedpiper.com}; and
(4)~the directory containing the site's files, which are copied from the puppet master
    server (indicated by the \texttt{puppet://} prefix).
The manifest also specifies three dependencies:
(1)~the \lstinline|apache2| package must be installed before the service;
(2)~the \lstinline|apache2| package must be installed before \lstinline|piedpiper.conf|
    is created; and
(3)~the \lstinline|apache2| service is ``notified'' (restarted) when
    \lstinline|piedpiper.conf|
    changes.

This manifest will successfully deploy, but that does not guarantee
that the resulting system configuration is correct. In fact, the manifest 
in \Cref{example1} has a
bug: if we visit \url{www.piedpiper.com}, we will get an HTTP 403
Forbidden error.

\myparagraph{Repairing the configuration in the shell}
When the system administrator discovers this problem, she considers that 
a 403 error may either indicate that the client does not have permission
to access the requested resource or that the server is misconfigured
and cannot access a needed file. For security reasons, the server does not
send a detailed error message to the client. Therefore, the
only way to debug the problem is to inspect the web server log. The
administrator runs the following command:
\begin{shellcode}
tail /var/log/apache2/error.log
\end{shellcode}
The log contains the line ``{\small\texttt{(13) permission denied}}'', which
indicates that the permissions on the site directory may be incorrect.
To investigate this, the administrator now runs the following command:
\begin{shellcode}
stat /var/sites/piedpiper
\end{shellcode}
The result of this command returns the directory's owner, which is \texttt{root}, and
its permissions, which is \texttt{0700}. This indicates that the directory is not readable
by others, including website visitors. The fix is to make the directory
readable by all:
\begin{shellcode}
chmod 755 /var/sites/piedpiper
\end{shellcode}
Now, the administrator can refresh the page and verify that the error is fixed.

Unfortunately, the state of the machine has now \emph{drifted} from
state specified in the manifest. Using this manifest to configure a second web
server will lead to the same problem and the administrator will have to manually apply
the same fix.  Worse, Puppet itself will undo the fix on this server! 
Any changes to the manifest, e.g., to install more software, will cause
Puppet to re-apply all resources and revert the permissions back to
their original broken state. When using Puppet, it is safe to perform
read-only actions to explore the machine state using the shell, e.g., to view
logs or inspect permissions. However, it is unsafe to use the shell to perform
updates.

\myparagraph{\sys solution}
\sys allows the administrator to fix the bug using the shell without the risk
of configuration drift. To do so, \sys first translates the manifest into
a program, written in \lang, that models all the effects that the manifest
has on the file system. The model is lengthy, but only a small fragment
is relevant to this repair:
\begin{lstlisting}[language=up,style=figurestyle]
rlet title = "/etc/sites/piedpiper" from str;
rlet mode = 0700 from int(9);
...
chmod(title,mode);
\end{lstlisting}
The code uses the \lstinline|chmod| command to set the mode.
However, instead of using constants for the mode and
the directory name, the command refers to the \emph{repairable variables} on
the first two lines. Each repairable variable specifies an original value and
a \emph{repair space} of alternate values. For example, on line 2,
the original mode is \lstinline|0700| but can be repaired to any
9-bit integer, if needed.
After producing this model, \sys translates the observed system calls issued
by the shell into an assertion, also expressed in \lang. In this case,
the assertion is as follows:
\begin{lstlisting}[firstnumber=5,language=up,style=figurestyle]
assert(mode("/var/sites/piedpiper") == 0755);
\end{lstlisting}
The only way for this assertion to hold is if the value of \lstinline|mode|
is repaired to \lstinline|0755| and the value of \lstinline|title| is left
unchanged. This repair to the \lang model
corresponds to changing line~\ref{line:permissions} of \Cref{example2}
to \lstinline|mode => 0755|.
This is exactly
the change the administrator would have made herself. However, in more sophisticated
manifests, there may be several alternative repairs that \sys ranks and presents
to the user as patches.

\begin{figure}[t]
\begin{centerbox}
\begin{puppetcode}
package{"apache2": ensure => present }
service{"apache2": ensure => running }

define website($title,$root) {
  file{"/etc/apache2/sites-enabled/$title.conf":
    content => "
    <VirtualHost $title:80>
    DocumentRoot /var/sites/$root
    </VirtualHost>" }
  
  file{"/var/sites/$root":
    ensure => "directory",
    source => "puppet://sites/$root",
    owner => root,
    mode => 0700, #\label{line-mode2}#
    recurse => "remote" } }

website{"www.piedpiper.com": root => "piedpiper" }#\label{line-piedpier}#
website{"piperchat.com": root => "piperchat" }#\label{line-piperchat}#
\end{puppetcode}
\end{centerbox}

\caption{A website abstraction. For exposition, the inter-resource
dependencies are elided.}
\label{example2}
\end{figure}

\myparagraph{User-defined abstractions in Puppet} We now consider a more
sophisticated example manifest that uses Puppet's abstractions to manage a
second website, \url{piperchat.com}.
The na\"ive approach is to
duplicate and tweak the configuration for \url{piedpiper.com}.
But, a
better approach is to create a custom \lstinline|website| type
(known as a \emph{defined type} in Puppet) for managing a website that is parameterized
by the domain name and site directory. This custom type allows websites
to be configured with just one line each (lines~\ref{line-piedpier} and \ref{line-piperchat} in \Cref{example2}).

Suppose the system administrator built this abstraction before fixing the
permissions problem, thus both websites produce the same error.
She discovers the error by visiting \url{piedpiper.com},
as she did earlier, witnesses the 403 error, and then uses the
shell to diagnose and fix the problem as before, using
the \lstinline|chmod| command.
Without \sys, there are two problems: (1)~the configuration has drifted from
the manifest as before and (2)~the other website remains broken.

\myparagraph{\sys solution}
There are two ways to correct the manifest to be consistent with the system
state:
\begin{enumerate}

\item Change line~\ref{line-mode2} to be consistent with the shell and affect
  both websites:
  \begin{puppetcode}
    mode => 0755
  \end{puppetcode}

\item Change line~\ref{line-mode2} to be exactly consistent with the shell
  and leave the other website unaffected:
  
  \begin{puppetcode}
  $title == "piperchat" ? 0755 : 0700
  \end{puppetcode}

\end{enumerate}
There are situations where either type of repair may be desired. The
first repair generalizes a change made to one instance to all
instances of the same type, whereas the second kind of repair is necessary
to specify special-case behavior.
In general, \sys cannot know which kind of repair
is desired, so it presents both repairs to the administrator. Since special-cases
are the exception, \sys ranks the repairs in the order shown above.
Notice that both repairs
update a line of code that is within a defined type, so \sys is not limited to
working with Puppet's built-in abstractions.

\begin{figure}[t]
\begin{centerbox}
\begin{puppetcode}
define website($title,$root,$https = false) {
  if ($https) { ... lots of configuration ... }
  else { ... same as before ... } }
\end{puppetcode}
\end{centerbox}
\caption{Optional support for HTTPS.}
\label{example3}
\end{figure}

\myparagraph{Reusing abstractions} 
\sys can also repair resources that instantiate user-defined types, as the next example
shows.
Suppose the administrator wants to start using HTTPS to secure websites.
Modern web browsers block
HTTPS servers from loading JavaScript
from unsecured domains. Therefore, sites need to be carefully upgraded
to ensure that all third-party code
is served over HTTPS too. For this reason, it makes sense to migrate
one website at a time.

The process of upgrading \url{www.piedpiper.com} to HTTPS
involves specifying the certificate, private chain, ciphers, and several
other details that are difficult to get right the first time.
Moreover, there is a risk that a faulty edit to the \lstinline|website| type will
inadvertently break the other website too. Therefore, it is safer to directly
edit the Apache configuration file for \url{www.piedpiper.com} instead.
Apache has a command-line tool to catch syntax
errors (\texttt{apachectl configtest}) that the administrator may use
for this task.
Once the server is working correctly, the administrator
can abstract the changes to make it easier to migrate other servers by
adding an optional \lstinline|https|
parameter to the \lstinline|website| type, as sketched in
\Cref{example3}.

However, the configuration has again drifted from the manifest. In this case,
\sys detects that
the configuration for \url{piedpiper.com} is a concrete instance
of the abstraction and automatically adds the
\lstinline|https| attribute:
\begin{puppetcode}
website{"www.piedpiper.com": root => "piedpiper", https => true }
\end{puppetcode}
This repair is notable because it repairs an instantiation of a type
that is not built-in to Puppet.

\myparagraph{Summary} We have presented three ways in which
\sys allows a system administrator to use Puppet and a shell in
harmony, benefiting from the unique strengths of each tool.  In all cases,
\sys synthesizes maintainable patches and ensures that no
configuration drift occurs.

\section{The \sys Repair Space}
\label{repair-space}

Puppet's own linting tools can help administrators fix syntax errors and type
errors. However, there are three more ways in which a manifest may need to
change. \sys helps administrators make the third type of change listed below.

\begin{enumerate}

\item \textbf{Adding, removing, or modifying dependencies.} The dependencies in a
  manifest impose a partial ordering on resources. Although Puppet automatically inserts
  certain dependencies (\emph{``auto-requires''}), others need to be
  specified explicitly by the administrator. Missing dependencies can cause a manifest to
  raise an error during deployment. \sys does not correct dependency
  errors, but this is the subject our prior work~\cite{shambaugh:rehearsal}.

\item \textbf{Creating new abstractions.} A powerful feature of Puppet is its
  ability to create new abstractions (defined types and classes) to make
  manifests modular and reusable. For example, in \Cref{motivation}, we
  created a \lstinline|website| abstraction to help manage a website.
  \sys does not help the user create new abstractions. However, given a manifest
  that has user-defined abstractions, \sys can perform repairs within them.

\item \textbf{Creating, deleting, and updating resources.}  
  \sys supports repairs that involve deleting resources and creating
  new resources, including instances of user-defined abstractions. In addition,
  \sys supports repairs that involve creating, deleting, and modifying
  attributes of existing resources, as detailed next.
\end{enumerate}

The rest of this section gives examples of \sys-supported repairs. We describe
individual repairs in isolation, but a single repair may involve
several repairs of the kind illustrated below.

\subsection{Supported Repairs}

\begin{figure}[t]
\begin{centerbox}
\begin{puppetcode}
file{"/fileA": content => "test"} #\label{line-filea}#

define T($x,$prefix) {
  if ($prefix) {
    file{"/dir/$x": content => "test"} } #\label{line-prefix-case}#
  else {
    file{$x: content => "test"} } }

T{x => "fileB", prefix => true}
T{x => "fileC", prefix => true}
\end{puppetcode}
\end{centerbox}
\caption{Repair example.}
\label{repair-space-example}
\end{figure}

The most significant class of repairs that \sys performs involves adding,
removing, and updating attributes on existing resources. Puppet has
dozens of resource types and each type has several
attributes that can dramatically change how the resource is interpreted.
What makes \sys powerful is its ability to correct the
attributes of both built-in and user-defined resources.  To
illustrate this, we use the manifest in \Cref{repair-space-example}, which
has one defined type, \lstinline|T|, and creates three files.  It
creates \lstinline|/fileA| directly, but uses the type \lstinline|T|
to create \lstinline|/dir/fileB| and \lstinline|/dir/fileC|.  An
interesting feature of \lstinline|T| is that it checks to see if the
\lstinline|$prefix| attribute is set, and if it is, it builds a
filename using string interpolation.

\myparagraph{Add new attribute}
\sys can add new constant-valued attributes to a resource. For
example, in \Cref{repair-space-example}, if the administrator uses the
shell to change the owner of \lstinline|/fileA| to
\texttt{alice}, \sys will add the attribute \lstinline|owner => alice|
to the corresponding resource.  If she instead changes the owner of
\lstinline|/dir/fileB| to \lstinline|alice|,
then \sys
suggests two possible changes, in order:
\begin{enumerate}

\item Add an attribute on line~\ref{line-prefix-case} that
  affects \lstinline|fileC| too:
  \begin{puppetcode}
  owner => alice
  \end{puppetcode}

\item Change line~\ref{line-prefix-case} to create a special case for
  \lstinline|fileB|, which does not affect \lstinline|fileC|:
  \begin{puppetcode}
  $title == "fileB" ? owner => alice : owner => root
  \end{puppetcode}

\end{enumerate}
\vspace{-1em}

\myparagraph{Delete existing attribute}
\sys can also delete attributes. For example, if the administrator 
changes the owner of \lstinline|/fileA| back to \lstinline|root|,
it will suggest removing the \lstinline|owner| attribute.

\myparagraph{Update existing constant}
In \Cref{motivation}, we saw that \sys can update the value of constants in
attributes. The same mechanism allows \sys to update attribute titles.
For example, renaming \lstinline|/fileA| to \lstinline|/fileA2| causes
\sys to update the manifest to refer to the new file (Line~1 of \Cref{repair-space-example}).
A harder repair involves renaming the files that are created indirectly by the
defined type. For example,
we could rename \lstinline|/dir/fileB| in three ways:
\begin{enumerate}
\item If renaming the file part, e.g., to \lstinline|/dir/fileB2|, the repair is in the  instantiation of \lstinline|T| (line~9).

\item If renaming the directory part, e.g., to \lstinline|/dir2/fileB|, the repair is
  in the definition of \lstinline|T| (line~3).

\item If renaming both, e.g., to \lstinline|/dir2/fileB2|, the repair must affect both locations.
\end{enumerate}
\sys supports all three repairs.

\myparagraph{Create and delete resource}
\sys can create and delete resources. For example, if the user deletes
\lstinline|/fileA|, \sys suggests removing line~\ref{line-filea} from
the manifest.  On the other hand, if the user creates a new file
\lstinline|/dir/fileD| with the same content specified in the
definition of \lstinline|T|, \sys suggests two fixes: (1) create a
new \lstinline|file| or (2) create a new \lstinline|T|
resource.

\subsection{Repair Consistency}
\label{consistency}

A key property of  \sys is that it produces \emph{consistent} repairs: 
a repair is guaranteed to preserve all changes made using
the shell that are within the scope of \sys's system model.
\Cref{sec:semantics} presents this model in detail, but at a high-level, we
model certain essential properties of regular files and directories, such as
their contents and permissions. In contrast, \sys does not support repairs that
affect running processes or special files, such as the \texttt{/proc} file
system. For example, many changes to the \texttt{/proc} file system are lost
after reboot, but can be persisted by editing certain configuration files in
the \texttt{/etc} directory. These kinds of repairs are straightforward in
principle, but would require a lot of engineering.

\section{From Manifests and Shell Commands to \lang}
\label{sec:semantics}

\Cref{sec:The lang Modeling Language} introduces our modeling language
\lang that provides a uniform way to model the semantics of manifests, the
constraints generated from shell commands, and the space of possible repairs.
Sections~\ref{sec:Primitive Resources} and \ref{sec:Used-Defined Resources}
describe primitive and user-defined resources. 
\Cref{sec:Creating and Deleting Resources} describes how we model
repairs that create and delete resources.
Finally, \Cref{sec:From Shell
Commands to Constraints} details how manifests and shell commands are
translated to \lang. \Cref{formula-generation} will describe translating
\lang into formulae for an SMT solver. While it is possible to directly
translate manifests and shell commands into constraints, using \lang has two
advantages: (1)~it is much easier to model the semantics of manifests in
\lang since it has imperative file-system operations and (2)~we can
simplify \lang programs before generating constraints, which makes
constraint solving scalable.

\begin{figure}[t]
\figsize
\centering
\lstset{language=up,basicstyle=\figsize\ttfamily}
\begin{tabular}{@{}r@{~}c@{~}l@{~}l@{}}
\multicolumn{4}{@{}l}{\textbf{Atomic Expressions}} \\
$a$ & ::=    & \textit{str}      & String \\
    & $\mid$ & \textit{bool}     & Boolean \\
    & $\mid$ & \textit{n}        & Integer \\
    & $\mid$ & \lstinline|undef| & Undefined \\
    & $\mid$ & x                 & Variable reference \\[.25em]
\multicolumn{4}{@{}l}{\textbf{Expressions}} \\
$e$ & ::=    & a                      \\
    & $\mid$ & \lstinline|file?($a$)|     & Test if $a$ refers to a file \\
    & $\mid$ & \lstinline|dir?($a$)|      & Test if $a$ refers to a directory \\
    & $\mid$ & \lstinline|exists?($a$)|   & Test if $a$ refers to a file or directory \\
    & $\mid$ & \lstinline|defined?($a$)|  & Test if not \lstinline|undef| \\
    & $\mid$ & \lstinline|$e_1$ + $e_2$|  & String concatenation \\
    & $\mid$ & $\cdots$                   & Comparisons and boolean connectives \\
  
\multicolumn{4}{@{}l}{\textbf{Statements}} \\[.25em]
$c$ & ::=    & \lstinline|let $x$ = $e$| & Variable declaration \\
    & $\mid$ & \lstinline|if ($e$) $c_1$ else $c_2$|  & Conditional \\
    & $\mid$ & \lstinline|{ $c_1$; $\cdots$; $c_n$ }| & Block statement \\
    & $\mid$ & \lstinline|chmod($e_1$,$e_2$)|
             & Set permissions of $e_1$ to $e_2$ \\
    & $\mid$ & \lstinline|chown($e_1$,$e_2$)|
             & Set owner of $e_1$ to $e_2$ \\
    & $\mid$ & \lstinline|mkdir($e$)|
             & Create directory \\
    & $\mid$ & \lstinline|write($e_1$,$e_2$)|
             & Create file $e_1$ with contents $e_2$ \\
    & $\mid$ & $\cdots$
             & Other file system operations \\
    & $\mid$ & \lstinline|rlet $x$ = $a$ from $r$|
             & \textrm{Let $x$ be $a$, but can be repaired to $r$} \\
    & $\mid$ & \lstinline|assert($e$)|  & \textrm{Assertion} \\[.25em]
\multicolumn{4}{@{}l}{\textbf{Repair Spaces}} \\
$r$ & ::=    & \lstinline|[$a_1$; $\cdots$; $a_n$]| & Finite set of alternatives\\
    & $\mid$ & \lstinline|str| & Any string or \lstinline|undef| \\
    & $\mid$ & \lstinline|int($n$)| & Any $n$-bit number or \lstinline|undef|
\end{tabular}

\caption{\lang Syntax}
\label{corelang-syntax}
\end{figure}

\subsection{The \lang Modeling Language}
\label{sec:The lang Modeling Language}

\lstset{language=up}\lang is an imperative language with primitive
operations that manipulate files, so it allows us to model
the side-effects that resources have on system state. In addition, it has two features that facilitate repair:
(1)~it has \emph{repairable variable declarations}, which are ordinary
variables that are augmented with a \emph{repair space} of alternate
values and (2)~it has assertions, which we use to constrain repair
spaces.  Intuitively, a single \lang program with repairable variables
represents a space of possible programs, ranked by the number and
kinds of repairs made.  The key to our approach is to translate
manifests to \lang programs with repairable variables and to turn
shell commands into \lang assertions that rule out
programs that are not consistent with the user's repair.

\myparagraph{File system operations}
\Cref{corelang-syntax} shows the syntax of \lang, which consists of
statements, expressions, atomic expressions, and repair spaces.
Atomic expressions include constants and variable references, which
are the simplest kinds of expressions that can appear in
manifests. Atomic expressions also include the special value \lstinline|undef|,
which we use to explicitly indicate that an optional attribute is not
present.

\lang's expressions include predicates to test if a path
refers to a file (\lstinline|file?|), a directory (\lstinline|dir?|),
or is non-existant (\lstinline|exists?|). 
These predicates only read file system state and do not perform writes.
For convenience,  \lang also has a predicate to test that an expression is not
the special value \lstinline|undef| (\lstinline|defined?|).
Finally, expressions include all atomic expressions as well as
conventional comparisons and boolean operators, which we elide from the figure.

\lang's statements have imperative operations that model
file system updates, including operations to create files
(\lstinline|write|), create directories (\lstinline|mkdir|), set file
permissions (\lstinline|chmod|), set file ownership
(\lstinline|chown|), and so on. \lang also has conditionals (\lstinline|if|),
immutable variables (\lstinline|let|), and block statements.

\myparagraph{Assertions and repairable variable declarations}
An unusual feature of \lang is that it supports \emph{repairable
  variable declarations}.  The statement
\lstinline|rlet $x$ = $a$ from $r$| binds the variable $x$ to the atomic
expression $a$ and
specifies that $r$ is its \emph{repair space}. \lang supports three
sorts of repair spaces:
\begin{enumerate}
\item A finite set of atomic expressions, which may include variables
   (\lstinline|[$a_1$; $\cdots$; $a_n$]|);
  \item The space of all strings (\lstinline|str|); and
  \item The space of $n$-bit integers, for a fixed $n$ (\lstinline|int($n$)|).
\end{enumerate}
A repairable variable also expresses the soft constraint that $x$ should be
$a$ if possible, thus there is a cost associated with picking an alternate
value from $r$.
In contrast,
an assertion expresses a hard constraint that cannot be
violated (\lstinline|assert($e$)|). One way to rank repairs would be
by the number of soft constraints violated, but \Cref{formula-generation} presents a more subtle
ranking procedure that works better in practice.

\subsection{Primitive Resources}
\label{sec:Primitive Resources}

We now present our  model of two key Puppet types.

\begin{figure}[t]
\begin{minipage}{0.4\columnwidth}
\begin{subfigure}{\columnwidth}
\begin{upcode}
let title = "/fileA";
let content = "Hello world";
if (exists?(title)) {
  rm(title);
}
write(title, content)
\end{upcode}
\caption{A trivial encoding.}
\label{encoding-trivial}
\end{subfigure}

\begin{subfigure}{\columnwidth}
\begin{puppetcode}
file{"/fileA"#\textsuperscript{t}#:
  content => "Hello world"#\textsuperscript{c}#
  source => undefined#\textsuperscript{s}#
  mode => undefined#\textsuperscript{m}#
  ensure => undefined#\textsuperscript{e}#
}
\end{puppetcode}
\caption{The annotated manifest.}
\label{annotated-file}
\end{subfigure}
\end{minipage}
\quad\begin{minipage}{0.45\columnwidth}
\begin{subfigure}{\columnwidth}
\begin{upcode}
rlet t = "/fileA" from str;
rlet c = "Hello world" from str;
rlet s = undef from str;
rlet e = undef from str;
rlet m = undef from int(9);
if (exists?(t)) { rm(t); }
if (e == "directory") {
  assert(c == undef and s == undef);
  mkdir(t)
}
else if (e == "file" or e == undef) {
  assert(defined?(s) xor defined?(c));
  if (defined?(s)) { cp(s, t); }
  if (defined?(c)) { write(t, c); }
}
else {
  assert(false);
}
if defined?(m) { chmod(t, m); }
\end{upcode}
\caption{A repairable encoding.}
\label{encoding-full}
\end{subfigure}
\end{minipage}

\caption{A file resource and a portion of its repair space.}
\end{figure}

\myparagraph{The file type}
The \lstinline|file| type only manages a single file, but it has
32 optional attributes, some of which dramatically alter its semantics.
For brevity, we only discuss five representative attributes, but our
implementation supports other attributes too:
\begin{enumerate}
\item The \lstinline|ensure| attribute determines if the resource is a
  file or directory. If omitted, it is assumed to be a file.
\item The \lstinline|content| attribute specifies the file source inline
  and the \lstinline|source| attribute copies contents from another file.
  These attributes are mutually exclusive. If the resource is managing
  a directory then neither may be defined.
\item The \lstinline|mode| attribute sets the file's permissions.
\end{enumerate}

\lang has the file system operations needed to model all the behaviors
described above. For example, consider the following resource which only specifies a single attribute:
\begin{puppetcode}
file{"/fileA": content => "Hello, world" }
\end{puppetcode}
We could model this resource as a trivial \lang program that deletes an
existing file or directory, if needed, and replaces it with the
specified file (\Cref{encoding-trivial}).\footnote{In practice, Puppet
  would not replace the file if it already had the specified
  contents. However, our simplified model is adequate for modeling
  repairs.} However, the resource needs to use repairable variables to
support repair.

To encode the full repair space, we take the following steps: (1)~we
produce a program with five repairable variables, one for the title
and four for each possible attribute (\Cref{encoding-full}); (2)~we
add all unused attributes to the resource and explicitly mark them
as undefined (\Cref{annotated-file}), and (3)~we annotate atomic
expressions in this manifest with the names of repairable variables.
The program in \Cref{encoding-full} first declares the repairable
variables, though note that all variables except \lstinline|c| and \lstinline|t|
are set to \lstinline|undef|. After these variables are declared, the program has
several cases that describe the space of all behaviors for a
\lstinline|file| resource. With no repairs, the program 
reduces to the trivial program in \Cref{encoding-trivial}.
However, repairs can make the other cases relevant.

For example, suppose the user removes the file and creates a directory with the
same name. This change produces the assertion
\lstinline|assert(dir?("/fileA"))|, which must hold at the end of the program.
The only way to satisfy this assertion, is to make the two following repairs:
(1)~the variable \lstinline|e| must be repaired to \lstinline|"directory"|,
since that is the only way that the branch with the \lstinline|mkdir| statement
is reachable, and (2)~the variable \lstinline|c| must be repaired to
\lstinline|undef|, since the branch asserts
that \lstinline|c| must be \lstinline|undef|. Finally, it is easy to
propagate the repair back to the manifest, since we had annotated atomic expressions
with their corresponded repairable variables.

\begin{figure}[t]
\begin{subfigure}[b]{0.48\columnwidth}
\begin{centerbox}
\begin{puppetcode}
package{"vim"#\textsuperscript{p}#:
  ensure => present#\textsuperscript{e}#
}
\end{puppetcode}
\end{centerbox}
\caption{The resource.}
\label{package-manifest}
\end{subfigure}
\vrule
\begin{subfigure}[b]{0.48\columnwidth}
\begin{centerbox}
\begin{upcode}
rlet p = "vim" from str;
rlet e = "present" from str;
rlet s = "dpkg" from str;
if (e == "present") {
  create(s + "://" + p, "");
}
else if (e == "absent") {
  rm(s + "://" + p);
}
\end{upcode}
\end{centerbox}
\caption{The \lang model.}
\label{package-model}
\end{subfigure}
\caption{A package resource and its model.}
\end{figure}

\myparagraph{The package type}
The \lstinline|package| type is very common in manifests and is a kind
of resource type that \sys models in a special way. We model a
resource that installs a package \texttt{p} using provider \texttt{s}
as a \lang program that creates an empty file called \texttt{s://p}
(\Cref{package-manifest}).  Conversely, we model a resource that
removes a package \texttt{p} using provider \texttt{s} as a \lang
program that deletes the file \texttt{s://p}. Since a repair may
either remove an installed package or change the package that is
installed, we translate a package resource into a \lang program with
three repairable variables: one for the title, one for the provider,
and one for the \lstinline|ensure| attribute, which determines if
package is present or absent (\Cref{package-model}).

To repair a package from the shell, the \sys user has to use standard
commands, e.g., \lstinline|apt install| or
\lstinline|apt remove|.\footnote{\texttt{apt} is the package manager
  on Debian-based systems. It should be straightforward to support other
  package managers too.}
When \sys monitors system calls from the shell, the system call trace
includes commands to launch these programs.
We translate invocations of these programs to constraints that create
and delete files in the \lstinline|dpkg://| path.
For example, the command \lstinline|apt remove vim| produces:
\vspace{-0.1em}
\begin{upcode}
assert(file?("dpkg://vim") == false)
\end{upcode}
This assertion does not hold after the program in \Cref{package-model}
executes, unless 
 we repair the variable  \lstinline|e|
to \lstinline|"absent"|. 

\myparagraph{Other types}
Puppet has several other built in types (48 as of this writing), many of which
are operating system-specific. With two exceptions, all types update
the state of the file system. Our implementation
supports several other common types, such as user accounts, SSH keys, cron jobs, and
more. \lang makes it easy to add support for new types, since it has the primitives
needed to model types and their repair spaces. The only two resource types
that do not update the file system are (1) \emph{notify}, which prints a log
message and has no effect and (2) \emph{service}, which starts and stops
running services. The former type is irrelevant for repairs and the latter
could be supported with some extensions to \lang.

\subsection{User-Defined Resources}
\label{sec:Used-Defined Resources}

\begin{figure}[t]
\begin{subfigure}[b]{0.48\columnwidth}
\begin{centerbox}
\begin{puppetcode}
define T($title) {
  file{$title + "/A"#\textsuperscript{y}#:#\label{line-a}#
    content => "textA" }
  file{$title + "/B"#\textsuperscript{z}#:
    content => "textB" }
}
T{"/dir1"#\textsuperscript{x}#: }#\label{line-dir}#
\end{puppetcode}
\end{centerbox}
\subcaption{Defined type.}
\label{defined-type-original}
\end{subfigure}
\vrule
\begin{subfigure}[b]{0.48\columnwidth}
\begin{centerbox}
\begin{upcode}
rlet x = "/dir1" from str;
rlet y = "/A" from str;
rlet z = "/B" from str;
let title0 = x + y;
let title1 = x + z;
...
\end{upcode}
\end{centerbox}
\subcaption[c]{\lang model.}
\label{defined-type-model}
\end{subfigure}
\caption{A naive expansion of a defined type.}
\label{defined-type}
\end{figure}

A manifest can define new resource types, known as \emph{defined
  types}.  A defined type can be thought of as function that produces
a manifest. For example, the manifest
in \Cref{defined-type-original} defines a type \lstinline|T| that takes
a directory name as a parameter and produces two file resources within
that directory. The manifest uses \lstinline|T| to create two files in
the directory \lstinline|/dir1|. Suppose we use the shell to rename
the file \lstinline|/dir1/A| to \lstinline|/dir2/C|. The only way to make
this edit is to change \lstinline|dir1| to \lstinline|dir2| (line~\ref{line-dir})
and \lstinline|A| to \lstinline|C| (line~\ref{line-a}). The former edit
has the added effect of renaming \lstinline|/dir1/B| to \lstinline|/dir2/B|.
We express this dependency in the \lang model by never
duplicating atomic expressions in the manifest (\Cref{defined-type-model}).

\subsection{Creating and Deleting Resources}
\label{sec:Creating and Deleting Resources}

To support repairs that delete resources, we wrap the statements of each resource
in a conditional that is guarded by a repairable boolean variable with the initial value
\lstinline|true|. If the value of the boolean is repaired to \lstinline|false|, then
none of the resource's statements take effect, which corresponds to the resource
being deleted. We ascribe resource deletions a much higher cost than attribute edits.
We support repairs that create new resources in a similar way, by creating template
resources that are guarded with a repairable variable that is instead initialized
to \lstinline|false|.

\subsection{From Shell Commands to Constraints}
\label{sec:From Shell Commands to Constraints}

\sys does not parse
shell commands but instead intercepts all system calls made during repair.
 For
example, the system call \lstinline|mkdir("/dirA")| turns into the assertion
\lstinline|assert(dir?("/dirA"))|. In a single repair session, a user may make
and revert changes. For example, the command
\lstinline|rmdir /dirA| produces the assertion \lstinline|assert(!dir?("/dirA"))|.
However, if the user first creates and then removes the directory, simply joining
both assertions is contradictory. \sys handles this kind of case by
calculating the strongest postcondition of the system call sequence instead of
naively turning each call into an assertion.

\section{From \lang to Logical Formulae}
\label{formula-generation}

We now discuss how we translate \lang programs into logical formulae
for an SMT solver, specifically Z3-str~\cite{zheng:z3str}.  The
formulae that we produce use the theories of bit-vectors and equalities
between concatenated strings and string variables. In our encoding,
each model returned by the solver can be interpreted as a combination
of a repair, which assigns values to the repairable variables, and a
set of variables indicating which repairable variables have changed
from their initial value.

\newcommand{\fs}{\oset{\rightharpoonup}{\mathit{fs}}}

At a high-level, we transform a \lang program into a formula ($\phi$)
with the following variables:
\begin{itemize}
  \item $\fs_{\mathit{in}}$ and $\fs_{\mathit{out}}$ are sets of
    variables that model the initial and final state of the
    file system;

  \item $\oset{\rightharpoonup}{x}$ are the values assigned to the repairable
    variables (whether or not they are repaired); and

  \item $n$ counts the number of repairs made.
\end{itemize}
We generate $\phi$ such that for all assignments to these variables,
$\phi$ is true, if and only if the modeled program updates the initial
file system ($\fs_{\mathit{in}}$) to the final file system
($\fs_{\mathit{out}}$), with exactly $n$ repairs to the repairable
variables ($\oset{\rightharpoonup}{x}$). Therefore, our goal is to
find an assignment to the repairable variables such that $\phi$ holds
for all input and output file systems:
\vspace{-0.5em}
\begin{mydisplaymath}
  \exists n,\oset{\rightharpoonup}{x} . \forall \fs_{\mathit{in}},\fs_{\mathit{out}} .
\phi\left(n,\oset{\rightharpoonup}{x},\fs_{\mathit{in}},\fs_{\mathit{out}}\right)
\end{mydisplaymath}
To produce solutions ordered by the number of repairs, we iteratively increase $n$
and search for $\oset{\rightharpoonup}{x}$ using counterexample-guided inductive
synthesis~\cite{solarlezama:sketch}.

\myparagraph{Encoding file systems}
We model each path ($p$) using four variables per path:
\begin{itemize}
  \item The state of the path ($s_p$): is $s_p$ a file, directory, or none;
  \item The contents ($c_p$), if $c_p$ is a file;
  \item The owner ($o_p$), if $o_p$ is a file or directory; and
  \item The mode ($m_p$), if $m_p$ is a file or directory.
\end{itemize}
We model a file system by modeling every possible path.  Although the
space of paths is potentially unbounded, we only need to consider the
(prefixes of) paths that appear in the \lang program. Recall that we
encode repairs as assertions, therefore we model all paths
that a repair affects, even if the repair affected paths that did not
appear in the original manifest.

\begin{figure}[t]
  \figsize
  \centering
  \begin{tabular}{l|l}
    \lstinline|exists?($p$)|
    & $s_p = \textbf{dir} \vee s_p = \textbf{file}$ \\[.25em]
    \lstinline|mode($p$) = 0700|
    & $s_p \ne \textbf{none} \wedge m_p = \textbf{0755}$ \\[.25em]
    \lstinline|contents($p$) = "hello"|
    & $s_p = \textbf{file} \wedge c_p = \texttt{"hello"}$
  \end{tabular}
  \caption{Examples of expressions and their encodings.}
  \label{encoding-examples}
\end{figure}
\myparagraph{Encoding expressions and statements}
Since \lang expressions only read the state of the file system,
they turn into predicates. \Cref{encoding-examples} translates some example 
expressions to predicates.
Since \lang statements update the state of the file system, we model
them as relations between two sets of variables that represent the
input and an output file system. For example, the statement \lstinline|mkdir(/x)|
constraints the state of \texttt{/x} in the output file system
($s'_\texttt{/x}$) to be a directory. The mode and owner are also set
to \lstinline|0755| and \lstinline|"root"| respectively, which are
Puppet's defaults. The content variable ($c'_\texttt{/x}$) is left
unconstrained, which is safe to do, since its value is uninterpreted
for directories. Finally, the relation constrains the variables for all
other paths such that they are the same in the input and output state:
\vspace{-0.5em}
\begin{mydisplaymath}
\begin{array}{l}  
s'_{\texttt{/x}} = \textbf{dir}~\wedge o'_{\texttt{/x}} = \texttt{"root"}~\wedge 
  m'_{\texttt{/x}} = \textbf{0755}~\wedge \\    
\forall p . p \ne \texttt{/x} \Rightarrow (s_p = s'_p \wedge c_p = c'_p \wedge 
     o_p = o'_p \wedge m_p = n'_p)
\end{array}
\end{mydisplaymath}
We translate all other primitive statements
in a similar way. Finally, we translate blocks and conditionals
by introducing intermediate states and flattening nested
conditionals.

\myparagraph{Encoding Repairable Variables}
A repairable variable, \lstinline|rlet $x$ = $a$ from $r$|, turns
into a new existentially quantified variable ($x$) with the specified domain
($r$). A repairable variable also has a cost, which is defined as follows: if
the value of the variable in a model is equal to the original value ($a$),
then the cost is $0$, otherwise the cost is $1$. The total cost of repairing
a manifest is the sum of all unit costs.

\myparagraph{Optimizing Update Synthesis}
To speed up repairs for large manifests, we use a minimization procedure
that turns repairable variables into constants when it is provably safe
to do so, by propagating information from the shell-based repair to the \lang model
of the manifest.
We transform the 
\lang program, translating repairable variable declarations for paths
not affected by the shell commands to ordinary let bindings. In doing
this, we have shrunk the overall number of repairable declarations
substantially, making the overall \sys performance based more around
the size of the update rather than the size of the manifest
 (\Cref{sec:Scalability Experiments}).

\myparagraph{Ranking Repairs}
Each model produced by the solver can be interpreted as a repair
and a ranking. 
\sys first ranks repairs by the number of repairable variables
changed, favoring repairs that makes fewer changes. To break ties
between repairs with the same number of changes, \sys favors repairs
that make fewer changes within defined types. The intuition behind
this tie-breaking procedure is that changes within defined types have
the potential to affect more resources, whereas changes outside defined
types only affect a single resource. Therefore, \sys primarily
ranks repairs based on the number of syntactic edits, but the
secondary ranking favors repairs typically make fewer semantic
changes. However, all repairs produced by \sys are consistent
with the changes made from the shell (\Cref{consistency}).

\myparagraph{Applying Updates}
Once a repair is chosen, applying an update is straightforward. Recall
that we annotate each atomic expressions in the manifest  with the name
of the repairable variable that holds its value. We only update
those atomic expressions whose repairable variables have
been updated.

\section{Evaluation}
\label{sec:evaluation}

We evaluate \sys in three ways. \Cref{sec:Evaluating Patch Rankings}
evaluates the quality of \sys-synthesized patches in an experiment by
measuring how highly \sys ranks correct patches on real-world manifests.
\Cref{sec:Applicability to Real-World Manifests} presents case studies of
some of these manifests. Finally, \Cref{sec:Scalability Experiments} evaluates
\sys's scalability on synthetic benchmarks.

\subsection{Evaluating Repair Rankings}
\label{sec:Evaluating Patch Rankings}

\begin{figure}[t]
  \figsize
  \begin{center}
  \begin{tabular}{lcccc@{}}    
\toprule
 \multirow{2}{*}{Benchmark} &    \# of     & \# of repair & \sys         & Average    \\
                            &    resources & scenarios    & runtime (ms) & repair rank\\
\midrule
  amavis & \phantom{00}6 & \phantom{0}1 &   \phantom{0,0}25 & 1.00\\
    bind & \phantom{00}6 & \phantom{0}3 &   \phantom{0,0}21 & 1.60\\
  clamav & \phantom{00}6 & \phantom{0}2 &   \phantom{0,0}23 & 3.50\\
 hosting & \phantom{0}19 & \phantom{0}1 &   \phantom{0,0}26 & 1.00\\
     irc & \phantom{0}18 & \phantom{0}1 &   \phantom{0,}292 & 1.00\\
     jpa & \phantom{0}10 & \phantom{0}1 &   \phantom{0,0}21 & 1.00\\
logstash & \phantom{0}14 & \phantom{0}6 &   \phantom{0,0}48 & 1.00\\
   monit & \phantom{00}7 & \phantom{0}4 &   \phantom{0,0}25 & 1.00\\
   nginx & \phantom{00}9 & \phantom{0}4 &   \phantom{0,0}27 & 1.00\\
     ntp & \phantom{00}4 & \phantom{0}3 &   \phantom{0,0}18 & 1.33\\
powerdns & \phantom{00}5 & \phantom{0}7 &   \phantom{0,0}39 & 1.43\\
 rsyslog & \phantom{00}7 & \phantom{0}4 &   \phantom{0,}129 & 1.25\\
  xinetd & \phantom{00}4 & \phantom{0}5 &             1,970 & 1.20\\
\midrule
Total    &           115 &           42 &   \phantom{0,}205 & 1.31\\
\bottomrule
  \end{tabular}
  \end{center}
  \caption{Benchmark of real-world Puppet manifests~\cite{shambaugh:rehearsal}.
  We identified a total of \scenecount scenarios in which the manifests would require
  repair. On average, \sys took 205 ms to compute the repairs, and the average
  rank of the ideal update was \avgrank.}
  \label{fig:github-summary}
\end{figure}

\begin{figure*}[t]
\begin{subfigure}[t]{0.32\textwidth}
\begin{puppetcode}
  $params::package_name = "pdns-server"
  $params::package_provider = "dpkg"
  define powerdns::install(
    $package = $params::package_name,
    $ensure = present,
    $source = undefined,
    ...) {
    package {$package: 
      ensure => $ensure,
      source => $source,
      provider = $params::package_provider,
    }
  }
powerdns::install { ensure => present }
\end{puppetcode}
\caption{PowerDNS}
\label{fig:powerdns-example}
\end{subfigure}
~\vrule~
\begin{subfigure}[t]{0.32\textwidth}    
\begin{puppetcode}
define ntp ($logfile = 'false', ...) {
  if ($logfile != 'false') {
    file { '/etc/logrotate.d/ntpd':
      ensure  => present,
      ...
    }
  }
  ...
}
ntp { logfile => true, ... }
\end{puppetcode}
\vspace{5.3ex}
\caption{NTP}
\label{ntp-example}
\end{subfigure}
~\vrule~
\begin{subfigure}[t]{0.32\textwidth}
\begin{puppetcode}
define xinetd($server_args, $port, ...) {  
  file { "/etc/xinetd.d/rsync":
    ensure  => present,
    content => "$server $server_args $port",
  }
}

$cf = '/etc/rsync.conf'#\label{xinetd-line}#
$args = "--daemon --config $cf"
xinetd {
  server_args => $args,
  port => 873
}
\end{puppetcode}
\vspace{1ex}
\caption{xinetd}
\label{xinetd-example}
\end{subfigure}
\caption{Portions of three manifests from our benchmarks.}
\end{figure*}

We studied the 13 manifests in a Puppet
benchmark~\cite{shambaugh:rehearsal} to identify scenarios in which
the manifests may need to be repaired. (\Cref{sec:Applicability to Real-World Manifests} highlights
some of these repairs.)  We identified a total of \scenecount such repair
scenarios. For each scenario, we used \sys to perform
\technique, instead of manually patching the manifest. We ran \sys 50 times
per scenario to measure average performance.
\Cref{fig:github-summary} summaries the benchmarks,  their size in number of resources, the number of
distinct scenarios for that benchmark, and the average time \sys took to
perform the repair.

For our experiment, we configured \sys to produce the five, highest-ranked repairs for
each repair scenario.  We took each ranked list, randomized its order, and
presented the repairs to one of the authors. The author (without knowing
\sys's ranked order) selected that repair that captured the intent of the
shell commands and labeled it ``correct''. (Recall that while all repairs
are guaranteed to be consistent, some may capture and generalize the intent
of the shell command better than others.) The average rank of the correct
repair in \sys's ranked lists was \avgrank.
Overall, the highest-ranked repair was the correct repair 76\% of the time.

\subsection{Repairs to Real-World Manifests}
\label{sec:Applicability to Real-World Manifests}

We describe three different types of repairs from our benchmarks as case studies
of \sys usage.

\myparagraph{Operating system update}
Different Linux distributions offer the same packages under different package
names. For example, the \emph{PowerDNS} benchmark
(\Cref{fig:powerdns-example}) installs the \texttt{pdns-server} package on
Debian, but fails on Red Hat where the package is called
\texttt{pdns}. Running \lstinline|yum install pdns| from the Red Hat system
shell fixes the problem; here, \sys's highest-ranked patch is the correct
patch. It is notable that the  benchmark does not directly
create the package. Instead, it has a global variable that's bound to the
package name and is used within a defined type.

\myparagraph{Updating optional resources}
Many reusable manifests provide optional features that the administrator
may want to turn on or off and \sys can help with these repairs.
For example, the NTP benchmark (\Cref{ntp-example}) has an abstraction
 that optionally creates a log file. Suppose that the log is initially
enabled, but that it subsequently needs to be removed (e.g., because of limited disk space
or the log is deemed unnecessary). \sys allows the
administrator to simply delete the log file from the shell. Its highest-ranked
repair changes the \lstinline|logfile| flag, which is the right way to
perform this update.

\myparagraph{Configuration file updates}
Manifests often use string interpolation to create
 configuration files from templates.
\Cref{xinetd-example} shows a fragment of a benchmark that creates a configuration
file in this way. We used a
text editor to update the configuration file, changing
\texttt{rsync.conf} to \texttt{piperchat.conf}, and \sys
updated the variable on line~\ref{xinetd-line}.

\begin{figure}[t]
\begin{subfigure}[b]{0.48\columnwidth}
\begin{tikzpicture}
\node{\pgfimage[width=\columnwidth]{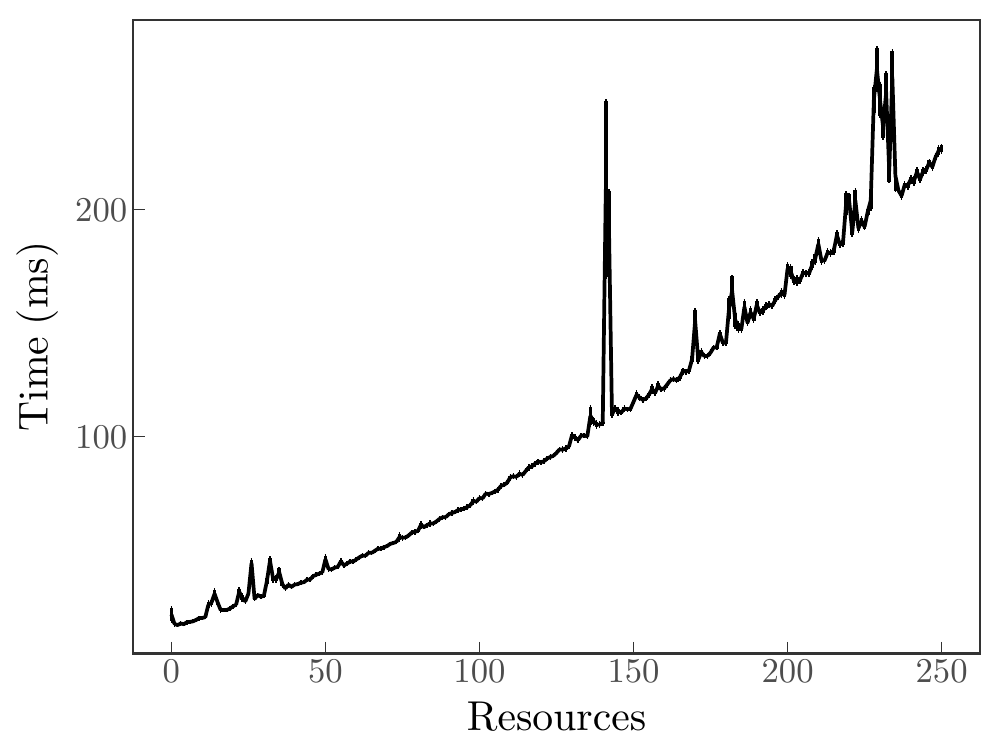}};
\end{tikzpicture}
\caption{Varying manifest size.}
\label{size-scaling}
\end{subfigure}
\begin{subfigure}[b]{0.48\columnwidth}
\begin{tikzpicture}
\node{\pgfimage[width=\columnwidth]{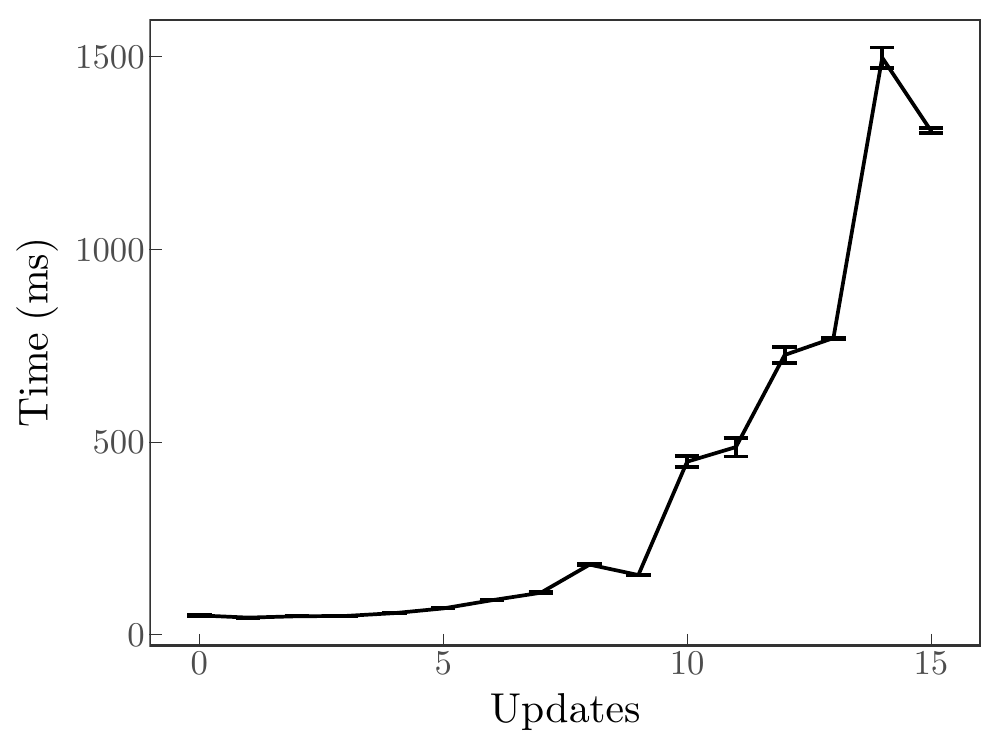}};
\end{tikzpicture}
\caption{Varying update size.}
\label{update-scaling}
\end{subfigure}
\caption{Scalability: average of 10 trials with standard error.}
\end{figure}

\subsection{Scalability Experiments}
\label{sec:Scalability Experiments}

\sys's running time is dominated by the SMT solver.
The size of each problem  depends
on (1)~the size of the manifest and (2)~the number of shell commands.
\Cref{size-scaling} shows how the
running time of \sys varies with the number of resources. We use a synthetic benchmark
that creates $n$ distinct resources and uses the same shell command to update
all manifests. The graph shows that \sys produces an update in less than 
1.5s, even when the manifest has 250 resources.
\Cref{update-scaling} shows how the running time of \sys varies with the
number of shell commands while the size of the manifest increases proportionally.
We generate a sequence of $n$ shell commands where each updates
a different resource. \sys takes up to 1.5s with 15 shell commands and
over a minute for more.

Although 15 shell commands may appear to be restrictive, note that we can
easily batch them. If a repair requires 30 shell commands, we can issue
the first 15 to get the manifest to an intermediate state and then issue the
next 15 to get the manifest to a final state. Also note that each shell command in the
benchmark is an update that leads to a repair. Shell commands that do not update
the file system do not generate constraints. Therefore, \sys allows the
administrator to explore the system as long as she likes.

\section{Scope and Limitations}
\label{sec:limitations}

\myparagraph{Threats to Experimental Validity}
For our evaluation, one author subjectively measured update correctness.
Real system administrators would provide a more accurate correctness
measure. We used a suite of benchmarks
collected from public GitHub repositories, but did not ask the
systems' developers to identify the repairs in these benchmarks. Instead,
we injected faults to create repair scenarios. A future user study of
industrial users could evaluate \sys's usefulness in practice.

\myparagraph{Unsupported Puppet Features}
Puppet is a sophisticated, evolving language and \sys supports a significant
subset of Puppet features. Our prototype does not support certain features
such as inheritance (which Puppet documentation states
should be used ``very sparingly'') and
lambdas (a recent language feature not yet widely
used). Nevertheless, it would be possible to add support for these features with
more engineering effort. Puppet also has two notable extra-linguistic features:
 manifests may have embedded shell scripts
(the \texttt{exec} type) and string templates written in Ruby (ERB). Repairing
these features are beyond the scope of this paper.

\myparagraph{Limitations of the \lang Model}
\Cref{repair-space} describes three classes of repairs, but \sys's
repair space only includes repairs that add, remove, or update
 resources. Therefore, \sys is
\emph{not complete} with respect to the full space of desirable repairs.
However, for its supported repairs, \sys produces repairs that are
\emph{consistent} with changes made from the shell that are within
the scope of \lang (\Cref{consistency}).
\lang only
models a few key attributes of regular files and directories. If a shell
command performs an update beyond the scope of the model, \sys will not detect
it. For example, if a manifest is configured to start a service and the user
terminates the service from the shell, \sys will not be able to repair the
manifest. Is is possible, in principle, to support this repair by enhancing
\lang to model processes and intercepting more system calls. In practice, \sys
will require careful engineering to support each kind of primitive resource.
This paper supports a subset of common primitive resources.

\myparagraph{Interaction Model}
During repair, we assume that all changes to the machine are made using
the \sys shell and we do not detect changes made by
background processes. In principle, it is straightforward to
support concurrent shells if their system call
logs can be totally ordered.

\section{Related Work}
\label{sec:related}

\myparagraph{Program Repair and Synthesis}
Fundamentally, Puppet manifests are programs and \sys is an automated
repair tool that uses shell commands as partial specifications of desired
behavior. This is not unlike most automated program repair 
tools~\cite{Weimer13, Jin11, Bradbury10, Arcuri08, Carzaniga13,
Wei10, Pei14, Liu12, Jeffrey09, Wilkerson12, Perkins09, Coker13, Debroy10,
Demsky06, Orlov11, Weimer09, LeGoues12a, LeGoues12b, Gopinath11, Carbin11,
Elkarablieh08, Qi15, Dallmeier09, Kim13, Tan15, nguyen:semfix, Sidiroglou05,
Qi13, mechtaev:directfix, mechtaev:angelix, Ke15ase, Sidiroglou-Douskos15,
Long15, Long16, DeMarco14, Jiang16} that use partial specifications, often
tests, to produce program patches that satisfy those specifications.

Our repair approach is a form of syntax-guided synthesis~\cite{alur:sygus}.
\sys models a space of possible repairs, similar to the repair models of Singh et
al.~\cite{singh:pythonfeedback}. Whereas they repair
student-written programs to conform to complete, teacher-provided
specifications, \sys uses partial specifications provided from shell
commands and ranks candidate repairs based on size.
\sys allows users to freely manipulate a manifest and its output while
propagating changes from one to the other, which is similar to
prodirect manipulation~\cite{chugh:prodirect}.

Synthesis-based program repair tools, e.g., 
Angelix~\cite{mechtaev:angelix}, DirectFix~\cite{mechtaev:directfix},
and SemFix~\cite{nguyen:semfix}, synthesize patches for more complex C programs than 
Puppet manifests. Because manifests are relatively simpler, \sys (1)~is 
much faster, (2)~generates and ranks multiple patches for the user to select the best one, and 
(3)~does not require the user to write tests, instead turning shell commands
into assertions to guide repair, which is a more natural interface
for system administrators.

\looseness-1
\myparagraph{Configuration Languages}
Automated testing and verification of system configuration languages has
focused on universal properties such as convergence~\cite{hanappi:convergence},
idempotence\cite{hummer:chef-idem}, and determinism~\cite{shambaugh:rehearsal}.
These universal properties are necessary, but insufficient for a manifest to be
correct. \sys is an interactive repair tool that can repair logic errors too.
ConfValley~\cite{huang:confvalley}, \textsc{PCheck}~\cite{xu:pcheck}, and
ConfigC~\cite{santolucito:configc} are complementary tools that validate
program-specific configuration files.

Tools like ConfSuggester~\cite{zhang:confsuggester}, AutoBash~\cite{autobash},
and ConfAid~\cite{confaid} find errors in configuration files, using dynamic
analysis to track how configuration values affect program execution. When a
Puppet manifest creates a buggy configuration file, it is the manifest the needs to
be repaired and not the generated configuration file itself. Given a fixed configuration,
\sys can repair a manifest and thus compliments these tools.

\looseness+1
$\mu$Puppet~\cite{fu:puppet-semantics} formalizes a subset of Puppet,
including many language features that \sys does not support. In contrast,
\sys models the effects that resources have on system state (i.e.,
 \emph{resource realization}), which is out of
scope for $\mu$Puppet.

\myparagraph{Shell Script Analysis} 
\sys complements shell script bug-finding tools, such as
\textsc{ABash}~\cite{mazurak:abash} and synthesis tools, such as
StriSynth~\cite{gulwani:strisynth}, as it works on Puppet manifests.

\myparagraph{Other configuration languages}
\sys leverages Puppet's DSL to model resources, which should be possible
for languages like Salt~\cite{salt},
Ansible~\cite{ansible}, and LCFG~\cite{anderson:lcfg}, but harder for
Chef~\cite{chef}, a Ruby-embedded domain-specific language.

\section{Contributions}
\label{sec:contributions}

\looseness+1
System configuration languages, such as Puppet, can make system administration
 easier. However,
manifests often have bugs and the shell is often the
best tool for diagnosing bugs.  Using \sys, administrators
can fix bugs using the shell, because \sys 
automatically synthesizes repairs to the underlying manifest. 
We have demonstrated that \sys is fast
on reasonably sized manifests, and that 76\% of the time, it
produces repairs equivalent to those written by humans. 

\section*{Acknowledgments}

We thank the anonymous reviewers,
Shriram Krishnamurthi, and Christian K\"{a}stner for their thoughtful
feedback. We thank Rachit Nigam for his work on an early
version of \sys.
This work is supported by the National Science Foundation under grants
CNS-1413985, 
CCF-1453474, 
CCF-1564162, 
CCF-1717636, 
and CNS-1744471. 
Aaron Weiss was affiliated with UMass Amherst during his primary work on \sys.

\balance

\bibliographystyle{plain}
\bibliography{paper}

\end{document}